\documentstyle[12pt]{article}
\newcommand{\newsubsection}[1]{
\vspace{1cm}
\pagebreak[3]
\addtocounter{subsection}{1}
\addcontentsline{toc}{subsection}{\protect
\numberline{\arabic{section}.\arabic{subsection}}{#1}}
\noindent{ \sc  #1}
\nopagebreak
\vspace{2mm}
\nopagebreak}
\newlength{\extraspace}
\newlength{\extraspaces}
\newcommand{\be}{\begin{equation}
\addtolength{\abovedisplayskip}{\extraspaces}
\addtolength{\belowdisplayskip}{\extraspaces}
\addtolength{\abovedisplayshortskip}{\extraspace}
\addtolength{\belowdisplayshortskip}{\extraspace}}
\newcommand{\ee}{\end{equation}}
\newcommand{\ba}{\begin{eqnarray}
\addtolength{\abovedisplayskip}{\extraspaces}
\addtolength{\belowdisplayskip}{\extraspaces}
\addtolength{\abovedisplayshortskip}{\extraspace}
\addtolength{\belowdisplayshortskip}{\extraspace}}
\newcommand{\ea}{\end{eqnarray}}
\newcommand{\nonu}{\nonumber \\[2mm]}
\newcommand{\is}{& \!\! = \!\! &}
\newcommand{\twomatrixd}[4]{{\left(\begin{array}{cc}
\displaystyle #1 & \displaystyle #2\\[2mm]
\displaystyle  #3  & \displaystyle #4 \end{array}\right)}}

\newcommand{\cP}{{\cal P}}
\newcommand{\hf}{{\textstyle{1\over 2}}}
\newcommand{\tr}{{\rm tr}}

\newcommand{\half}{{\textstyle{1\over 2}}}

\newcommand{\Z}{{\bf Z}}
\newcommand{\cH}{{\cal H }}
\newcommand{\cZ}{{\cal Z }}

\newcommand{\delbar}{\overline{\partial}}
\newcommand{\zbar}{{\overline{z}}}

\newcommand{\Qbar}{{\overline{Q}}}
\newcommand{\Lbar}{{\overline{L}}}

\def\a{\alpha}
\def\b{\beta}
\newcommand{\del}{\partial}
\def\G{\Gamma}
\def\e{\epsilon}

\newcommand{\cQ}{{\cal Q}}
\begin{document}
\addtolength{\baselineskip}{.5mm}
\newcommand{\dslash}{D}
\newcommand{\GG}{G}
\newcommand{\BPS}{{\mbox{\small\sc BPS}}}
\newcommand{\bps}{_{{}_{\rm BPS}}}
\newcommand{\LL}{_{\! {}_{L}}}
\newcommand{\RR}{_{\! {}_{R}}}
\newcommand{\pslash}{{p \hspace{-5pt} \slash}}
\newcommand{\aA}{{\bf a}}
\newcommand{\Ppsi}{{\bf \psi}}
\newcommand{\LR}{_{{}_{\! L,R}}}
\newcommand{\dnul}{\d_{{}_0}}

\newcommand{\lL}{{\mbox{\small  \cal L }}}
\newcommand{\gG}{{\mbox{\footnotesize \cal  G\,}}}
\newcommand{\sS}{{\lambda}}
\newcommand{\sSbar}{{\overline\lambda}}
\newcommand{\Bps}{{\mbox{bps}}}
\newcommand{\xX}{x}
\newcommand{\Lup}{^{\! {}^{L}}}
\newcommand{\NP}{{N_P}}

\begin{flushright}
March 1996\\
{\sc cern-th}/96-74\\
{\sc pupt}-1604
\end{flushright}
\vspace{-.3cm}
\thispagestyle{empty}

\begin{center}
{\Large\sc{BPS Spectrum of the Five-Brane\\[7mm]
           and Black Hole Entropy}}\\[13mm]

{\sc Robbert Dijkgraaf}\\[2.5mm]
{\it Department of Mathematics}\\
{\it University of Amsterdam, 1018 TE Amsterdam}\\[7mm]
{\sc Erik Verlinde}\\[2.5mm]
{\it TH-Division, CERN, CH-1211 Geneva 23}\\[.1mm]
and\\[.1mm]
{\it Institute for Theoretical Physics}\\
{\it Universtity of Utrecht, 3508 TA Utrecht}\\[7mm]
and \\[3mm]
{\sc  Herman Verlinde}\\[2.5mm]
{\it Institute for Theoretical Physics}\\
{\it University of Amsterdam, 1018 XE Amsterdam} \\[.1mm]
and\\[.1mm]
{\it Joseph Henry Laboratories}\\
{\it Princeton University, Princeton, NJ 08544}\\[18mm]

{\sc Abstract}

\end{center}

\noindent
We propose a formulation of 11-dimensional M-theory in terms of
five-branes with closed strings on their world-volume. We use this
description to construct the complete spectrum of BPS states in
compactifications to six and five dimensions. We compute the
degeneracy for fixed charge and find it to be in accordance with
U-duality (which in our formulation is manifest in six dimensions)
and
the statistical entropy formula of the corresponding black hole. We
also briefly comment on the compactification to four dimensions.

\vfill

\newpage
\newsubsection{Introduction.}

One of the outstanding challenges that one faces in building a
successful theory of quantum gravity is to provide a microscopic
description of black holes that explains and reproduces the
Hawking-Bekenstein formula for the entropy \cite{bh}. This question
has been hard to answer within string theory because, until recently,
black holes only arose as particular classical solitons of the low
energy effective field theory and hence could only be studied in
string perturbation theory. This situation has changed quite
dramatically through the recent developments related to string
duality
\cite{hulltownsend,wittenm}. It has been discovered that black holes
and other RR-solitons have an exact description in string theory in
terms of D-branes \cite{polch}, and that their quantum states are
related by duality to elementary strings excitations. These facts
have
been used recently in \cite{strovaf} to give a microscopic
description
in terms of D-branes of certain five-dimensional extremal black holes
and to check that it reproduces the expected entropy formula (see
also
\cite{blackholes}, and for extensions to 4 dimension
\cite{4dblackholes}).
The central idea of this computation is a comparison of the
asymptotic
growth of the number of BPS states in string theory with a given
charge and the area or volume of the horizon of the corresponding
black hole geometry. Specifically, the statistical entropy $S(Z)$ as
a
function of the charge $Z$ is expected to behave as
\be
S(Z) =   2\pi |Z|^\a
\ee
with $\a=2, {3\over 2}, 1$ for dimension $d=4,5,6$ respectively.
The precise form of the right-hand side is further restricted by
the required invariance under the U-duality symmetry group.

U-duality and other string dualities are still rather mysterious in
the D-brane description, because it necessarily makes a distinction
between charges that correspond to RR-fields and those of the
NS-sector of the theory. On the other hand, there are many
indications that
elementary strings and all other $p$-branes have a unified
description
in 11 dimensions \cite{townsend}. One therefore expects that
the U-duality invariance of the BPS spectrum can be explained
by extending the D-brane analysis to this eleven-dimensional M-theory
\cite{wittenm,schwarz,orbifolds}. It is known that M-theory contains
membranes and five-branes, which are charged relative to the
three-form gauge
field $C_3$ and its six-form dual $\tilde C_6$ respectively. However,
by forming bound states the five-brane can also carry $C_3$ charges
and
it therefore seems the natural starting point for a unified treatment
of
all BPS states in string theory.

It has been suggested that the five-brane can be viewed as
a D-object on which membranes can have boundaries \cite{mem}. In this
picture the world-volume theory of the five-brane is induced by the
boundary states of the membrane and thus is naturally described by a
{\it closed} string theory.  In this letter we propose to take this
description seriously.  We will present evidence that it indeed gives
rise to a complete, U-duality invariant counting of BPS states. In
our
presentation we will focus on the central ideas of the construction,
since the details of the calculations will be published elsewhere
\cite{inprep}.

\newsubsection{Strings on the Five-Brane}

For definiteness and simplicity, we will concentrate on {\it
toroidal}
compactifications of M-theory, because this will allow us to make
maximal use of the space-time supersymmetries. Also, we consider only
flat five-branes with the topology of $T^5$. The five-brane
represents
a soliton configuration that breaks half of the 32 space-time
supersymmetries of M-theory. On the $5+1$ dimensional world-volume
the
unbroken supercharges combine into 4 four-component chiral spinors
that generate a $N=(4,0)$ supersymmetry. It is known
\cite{callanetal}
that the {\it effective} world-brane theory is, after appropriate
gauge-fixing, described by a tensor multiplet containing $5$ scalars,
an anti-symmetric tensor $b^+$ with self-dual three-form field
strength
$db^+$, and 4 chiral fermions $\psi$. These fields must represent the
massless states of the closed strings that live on the world-brane.
One may think about these strings as solitons of the effective world
brane theory that are charged with respect to the tensor $b^+$. The
self-duality of $db^+$ implies that we are dealing with a
6-dimensional {\it self-dual} string theory (cf. \cite{sdstring}).

The string configurations break half of the world-brane
supersymmetries, and hence the world sheet formulation of this string
must contain 8 supercharges: 4 left-moving and 4 right-moving. The 8
remaining supercharges, those that are broken by the string, give
rise to $4$ left-moving and $4$ right-moving fermionic Goldstone
modes
$\sS$ and $\sSbar$. We assume that the world-sheet theory can be
formulated in a light-cone gauge, and so one expects to have $4$
bosonic fields $\xX$ that describe the transversal directions inside
the $5+1$ dimensional world-volume of the five-brane.  We find it
convenient to label these fields using chiral spinor indices $a$ and
$\dot a$ of the transversal $SO(4)$ rotation group.  In this notation
the left-moving fields on the string world-sheet are
\be
\label{fields}
\xX^{a\dot{a}}(z),\ \sS^a_\alpha(z),\
\ee
and the right-moving fields are
\be
\xX^{a\dot{a}}(\zbar),\ \sSbar_{\dot\alpha}^a(\zbar),
\ee
where we used the fact that the fermions $\sS$ and $\sSbar$ have the
same chirality with respect to the  $SO(4)$. The indices $\alpha$
($\dot\alpha$) are (anti-)chiral  spinor indices of another $SO(4)$
which, as will become clear,  becomes identified with part of the
space-time rotations. On the five-brane it will be realized as an
R-symmetry.

We propose to take the fields ($x^{a\dot a}, \sS^a_\a, \sSbar^a_{\dot
a}$) as the complete field content of the world-sheet theory in the
light-cone gauge, without worrying about a possible covariant
formulation. In fact, we have essentially half of the world-sheet
fields of the type IIB string in the Green-Schwarz
formulation\footnote{A covariant formulation of this string is likely
to have a gravitational anomaly, but this may be canceled by adding a
term of the form $\int b^+\tr R^2$ to the world volume action of the
five brane and by accompanying the world-sheet reparametrizations by
a gauge
transformation of the $b^+$ field. A similar mechanism is described
in the last reference of \cite{orbifolds}.} and
we can use this analogy to check that we get indeed the right
massless
fields. The ground states must form a multiplet of the left-moving
zero-mode algebra
$\{\sS_a^\alpha,\sS_b^\beta\}=\epsilon_{ab}\epsilon^{\alpha\beta}$.
This gives 2 left-moving bosonic ground states $|\alpha\rangle$ and 2
fermionic states $|a\rangle$. By taking the tensor product with the
right-moving vacua one obtains in total $16$ ground states
\be
\Bigl(|\a, k\LL\rangle \oplus |a,k\LL\rangle\Bigr)\otimes
\Bigl(|\dot\b, k\RR\rangle \oplus |b,k\RR\rangle\Bigr).
\ee
Here we also took into account the momenta $(k\LL, k\RR)$, which form
a $\Gamma_{5,5}$ lattice, since we have assumed that the five-brane
has the topology of $T^5$. Level matching implies that
for the ground states one should have that $|k\LL|=|k\RR|$. Notice
that these ground states are stable as long as the pair $(k\LL,k\RR)$
is a primitive vector on $\Gamma_{5,5}$.

Ignoring for a moment the string winding numbers, we find that the
ground states indeed represent the Fourier modes of the massless
tensor multiplet on the five-brane.  Specifically, the states
$|\alpha\dot\beta\rangle$ describe four scalars
$X^{\alpha\dot\beta}=\sigma_i^{\alpha\dot\beta} X^i$ that transform
as
a vector of the $SO(4)$ R-symmetry, while $|\alpha b\rangle$ and
$|a\dot{\beta}\rangle$ describe the eight helicity states of the 4
world-brane fermions $\psi^\alpha$ and $\psi^{\dot\alpha}$.  Finally,
the RR-like states $|ab\rangle$ decompose into a fifth scalar, which
we call $Y$, and the 3 helicity states of the tensor field $b^+$.
These are indeed the fields that parametrize the collective
excitations of the five-brane soliton. In our description, however,
these are just the low energy modes. The complete set of fluctuations
of the five-brane are parametrized by the quantum string states.  We
also note that from the point of view of the string world-sheet the 5
scalars on the five-brane that describe the transverse oscillations
naturally split up into four $X^i$'s plus one ``RR''-scalar $Y$.  The
interpretation of these fields in terms of the five-brane geometry
will be discussed in detail in \cite{inprep}.  {}From this point of
view it can be shown that $Y$ is naturally compactified.

Just as in the type II superstring one can combine the 4 left-moving
and 4 right-moving supercharges $G^{\dot{a}\a} =\oint
\del \xX^{\dot a b} \sS_b^\a$ and $\overline{G}^{\dot a \dot\a}=\oint
\delbar \xX^{\dot a b} \sSbar_b^{\dot\a}$ together with the eight
fermion zero-modes of $\sS^{a\a}$ and $\sSbar^{a\dot\b}$ to construct
the $N=(4,0)$ supercharges $\cQ$ on the world-volume of the
five-brane.  The fact that the string lives on a 5-torus, however,
has
important consequences for the supersymmetry algebra. Namely, the
anti-commutator of the left-moving supercharges produces the
left-moving momentum $k\LL$, while the right-movers give $k\RR$.
Hence the string states form representations of the $N=(4,0)$
supersymmetry algebra\footnote{We use a Hamiltonian notation with
only
the spatial rotation group $SO(5)$ manifest. Hence from now on the
indices $a$ and $\dot a$ are combined into a four-valued $SO(5)$
spinor index $a$. The index $m$ corresponds to a $SO(5)$ vector.}
\ba
\label{left}
\{ \cQ^{a\a} ,\cQ^{b\b}\} \is
\e^{\a\b}( \cP^0 {\bf 1}^{ab} + \cP\LL^m\Gamma_m^{ab}),
\nonu
\{ \cQ^{a\dot\a} ,\cQ^{b\dot\b}\} \is
\e^{\dot\a\dot\b}( \cP^0 {\bf 1}^{ab} + \cP\RR^m\Gamma_m^{ab}),
\ea
where $\Gamma_m^{ab}$ are $SO(5)$ gamma matrices. The operators
$\cP^0$, $\cP\LL^m$ and $\cP\RR^m$ act on multi-string states that
form the Hilbert space of the five-brane.  The five-brane Hamiltonian
$\cP^0$ measures the energy of the collection of strings, while
$\hf(\cP\LL^m +\cP\RR^m)$ measure the total momentum. But we see that
the algebra also contains a vector central charge $\cP\LL^{m}\! -\!
\cP\RR^{m}$, which measures the sum of the string winding numbers
around the 5 independent one-cycles on the $T^5$ of the world-brane.

\newsubsection{Counting Multiple BPS Strings}

The self-dual string is an interacting theory.
It has no weak coupling limit, since its coupling is fixed by
the self-duality relation. In the following, however, we will
assume that for the purpose of counting the number of BPS-states,
it will be an allowed procedure to treat it as a theory of
non-interacting strings. The BPS-restriction should indeed limit
the possible interactions that can take place. Furthermore, we will
find that our degeneracy formulas will be consistent with
previous results obtained from D-brane technology
\cite{vafa6d,sen,strovaf}, as well as with U-duality.

Our aim is to count BPS states of the space-time theory that respect
either $1/4$ or $1/8$ of the supersymmetries. On the
world-brane this translates to a condition that either $8$ or $4$ of
the 16 supercharges annihilate the states. The strongest BPS
condition is obtained by demanding that
\be
\label{bps}
\varepsilon_{a\a}\cQ^{a\a}|\BPS\rangle =0,
\ee
for four independent spinors $\epsilon_{a\alpha}$, and at the same
time imposing four similar relations for the supercharges
$\cQ^{a\dot\b}$.  By dropping these latter four relations one gets a
weaker BPS condition that, as we will see, allows many more states.
In fact, we can easily treat both cases in parallel.

Without loss of generality we can assume that $|\BPS\rangle$ is an
eigenstate of $\cP\LL^m$ and $\cP\RR^m$ with eigenvalue $P\LL^m$ and
$P\RR^m$. Since the condition (\ref{bps}) holds for all the states
$|\BPS\rangle$ in the same multiplet, we can use the algebra
(\ref{left}) to deduce that
\be
\label{blp}
\varepsilon_{\a a} ( P^0 {\bf 1}^{ab} + P\LL^m\Gamma_m^{ab}) =0.
\ee
The equation (\ref{blp}) only has solutions when $P^0=\pm|P\LL|$.
To count how many states we have for a given value of $P\LL^m$ we
have
to consider the multiple string states that have a total left-moving
momentum $P\LL^m\sim\sum k\LL^m$ and total energy $|P\LL|$.  Since
the
energy of a single string is bounded from below by $|k\LL|$,
we deduce that in fact all momenta $k\LL^m$ must be in the {\it
same} direction, namely that of $P\LL^m$.  For generic $\Gamma_{5,5}$
lattice this implies that $(k\LL,k\RR)$ must be a multiple $\ell$ of
the {\it primitive} vector $(\hat{P}\LL,\hat{P}\RR)\in
\Gamma_{5,5}$ in the direction of $(P\LL,P\RR)$. Thus we have
\be
(P\LL,P\RR)=\NP(\hat{P}\LL,\hat{P}\RR),
\ee
where $\NP$ is an integer which is defined by this equation.
Thus we see that in a BPS state of the five-brane, all world-brane
momenta and windings of the strings are in the same direction.
In other words, the BPS restriction implies that the five-brane
behaves effectively as a 1+1-dimensional string-like object, as
will become more evident in the following.

Now let $\cH_\ell$ denote the space of single string BPS states with
momentum $k=\ell\hat{P}$ that are in the left moving ground state.
Via the level matching condition, its dimension is given by (here
$\hat{P}^2=|\hat{P}\LL|^2-|\hat{P}\RR|^2$)
\be
\dim \cH_\ell = d(\hf \ell^2 \hat{P}^2),
\ee
which is the coefficient of the term $q^{\hf \ell^2 \hat{P}^2}$ in
the
elliptic genus of $T^4$
\be
\label{TT}
\sum_N d(N) q^N=16\prod_n{\left({1+q^n\over 1-q^n}\right)^4}.
\ee
We will now make the assumption that we can treat the second
quantized
theory on the five-brane as a theory of non-interacting strings.
The space of all multiple string states that satisfy the BPS
conditions,
and with a given total energy-momentum $P$, then takes the form
\be
\cH_P =\quad
{\mbox{\Huge$\oplus$}_{\strut
{}_{\!\!\!\!\!\!\!\!\!\!\!\!\!\!\!\!\!\!\!\!\!\!\!
\sum \ell N_\ell=N_P}}}
\mbox{\large $\otimes$}_{\strut{}_{\!\!\!\!\!\ell}}\
\ {\rm Sym}^{\strut N_\ell}  \cH_\ell,
\ee
where ${\rm Sym}^N$ indicates that $N$-th symmetric tensor product.
At first one may think that only the states with $\NP$ ``primitive''
strings with $\ell=1$ should contribute, because only these are
stable. This would imply that
only the first term ${\rm Sym}^{\NP} \cH_1$ must be kept. However, we
propose that the correct interpretation of the Hilbert spaces
$\cH_\ell$ with $\ell>1$ is that they represent the contributions of
the various fixed points of the permutation group, and hence describe
the bound states of $\ell$ different primitive strings.  These
therefore also
represent stable states, and should be counted as well. We thus
conjecture that the exact dimension of $\cH_P$ is determined by the
character expansion\footnote{Here we included a prefactor $(16)^2$,
which represents the dimension of the zero-mode Hilbert space on the
five-brane world-volume.}
\be
\label{dimsum}
\sum_{\NP}q^\NP \dim \cH_P = (16)^2 \prod_\ell \left(1+q^\ell\over
1-q^\ell\right)^{\hf d(\hf \ell^2\hat{P}^2)}.
\ee
This formula somewhat resembles
the expressions of Borcherds \cite{borcherds} for the denominator
formula of a generalized Kac-Moody algebra. Presumably, by a similar
calculation as in \cite{harveymoore} one can relate its logarithm
to the one-loop amplitude of the self-dual string.
We further note that for $\hat{P}^2=0$ (\ref{dimsum})
reduces to the standard (chiral) superstring partition function
(since $\hf d(0)=8$). This correspondence will be explained below.

An alternative but presumably equivalent description of the second
quantized BPS string states on the five-brane is obtained by
considering the sigma model on the ``target space'' $\sum_N {\rm
Sym}^NT^4$.  The intuitive picture behind this representation is that
the light-cone description of an $N$-string state may also be
thought of as
that of a single string state on the $N$-fold symmetric product of
its
target space.\footnote{This proposed representation of the
multi-string
Hilbert states was motivated by the construction described in
\cite{vafa6d,strovaf}, where the same sigma
model was used to encode all possible D-brane configurations.}
This description should be further supported by the
correspondence between the above formula for the dimension of $\cH_P$
and the term at order $q^{\hf \NP\hat{P}^2}$ in the expansion of the
elliptic genus of the orbifold ${\rm Sym}^{N_P} T^4$.  In this way of
looking at it one finds that the asymptotic growth equals that of
states at level $h=\hf N_P \hat{P}^2$
in a unitary conformal field theory with central charge $c=6\NP$.
The standard degeneracy formula gives
\be
\label{asymp}
\dim \cH_P \sim \exp
\Bigl(2\pi\sqrt{\hf(P\LL^2-P\RR^2)} \Bigr).
\ee
This same result can be obtained directly from (\ref{dimsum}).
We will now turn to the discussion of the
space-time meaning of the momentum vector $P$.

\newsubsection{U-Duality Invariant BPS Spectrum in $D=6$.}

Let us now focus on the space-time interpretation of these BPS states
in toroidal compactifications of M-theory. We consider situations
in which at least $5$ of the coordinates are compact so that the
five-brane can wrap completely around an internal $T^5$. We will also
assume that at least 4 coordinates are uncompactified because these
will be identified with the fields $X^i$ on the five-brane. This
leaves us with two cases: $d=6$ and $d=5$.

We start with $d=6$.  The space-time effective action for $M$
theory compactified on $T^5$ is given by $N=(4,4)$ six-dimensional
supergravity that contains as bosonic fields, besides the metric, 5
anti-symmetric tensors, 16 gauge fields and 25 scalars. The scalars
parametrize a $SO(5,5)/SO(5)\times SO(5)$ manifold and hence the
expected U-duality group is $SO(5,5,\Z)$.  The $N=(4,4)$
supersymmetry algebra is
\ba
\left\lbrace Q^a_\alpha,Q^b_\beta\right\rbrace\is \omega^{ab}
\pslash_{\alpha\beta},\nonu
\left\lbrace Q^a_\alpha,\Qbar^b_{\overline{\beta}}\right\rbrace \is
\delta_{\a\overline\b} Z^{ab},
\ea
where $a,b = 1,\ldots,4$ are now $SO(5)\cong Usp(4)$ spinor indices
and $\omega^{ab}$ is an anti-symmetric matrix, that will be used to
raise and lower indices. The algebra contains 16 central charges that
are combined in the $4\times 4$ matrix $Z^{ab}$, where $a,b =
1,\ldots,4$ are again $SO(5)$ spinor indices. The central charge
$Z^{ab}$ forms a 16 component spinor of the $SO(5,5,\Z)$ U-duality
group, and takes its values on an integral lattice whose shape is
determined by the expectation values of the scalars.

Now let us look at the BPS states in the six-dimensional space-time
theory that respect $1/4$ of the space-time supersymmetries. The mass
of these BPS state can be determined from $Z^{ab}$ by solving the
eigenvalue equations
\ba
(Z^\dagger Z)^a{}_b\, \varepsilon\LL^b\is m\bps^2
\varepsilon\LL^{a},\nonu
(Z Z^\dagger)^a{}_b\, \varepsilon\RR^b\is m\bps^2 \varepsilon\RR^{a}.
\ea
Each of these equations has two independent solutions, as can be seen
for example from the fact that the matrices $Z^\dagger Z$ and
$ZZ^\dagger$ must be of the form
\ba
\label{relations}
(Z^\dagger Z)^{ab}\is (m\bps^2\! -2|K\LL|) {\bf 1}^{ab} +
2K\LL^m\Gamma_m^{ab},\nonu
(Z Z^\dagger)^{ab}\is (m\bps^2 \! -2|K\RR|){\bf 1}^{ab} +
2K\RR^m\Gamma_m^{ab}.
\ea
Since $|K\LL| =|K\RR|$ the combination $(K\LL,K\RR)$ forms a null
vector of $SO(5,5,\Z)$. As we will show below, BPS states with fixed
$(K\LL,K\RR)$ correspond to those multi-string states on the
five-brane for which the total momentum and winding number vector
$(P\LL, P\RR)$ is equal to minus $(K\LL,K\RR)$.  Since this implies
that $P\LL^2-P\RR^2=0$, the corresponding BPS states are necessarily
made up from the string ground states with $|k\LL|=|k\RR|$.

To derive the relations (\ref{relations}) we must understand how the
space-time supersymmetry algebra, including its central charge, is
realized on the world-volume of the five-brane. We assume that the
world-brane theory is formulated in a light-cone gauge, so that  the
$SO(5,1)$ space-time Lorentz group is broken to the $SO(4)$ subgroup
of transverse rotations. On the world-brane this group becomes
identified with the R-symmetry, and so the previously introduced
spinor indices $\a,\dot\a$ indeed correspond to the two spin
representations of the space-time rotations.
Notice that the space-time chirality is thus linked with the
chirality on the string world-sheet.
To construct the $N=(4,4)$ space-time supersymmetry algebra on the
world-brane we need to use the zero-mode algebra
of the various world-brane fields.
The transversal momentum $p^{\a\dot \b}$ is as usual identified with
the zero-modes of the canonical conjugate of the fields
$X^{\a\dot\b}$.
Similarly, there are fermion zero-modes  $S^\a_b$ and $S^{\dot \b}_a$
that represent the broken part of the space-time supercharges. They
can be normalized such that they satisfy the algebra
\be
\{S^\a_a,S^\b_b\} = \e^{\a\b}\omega_{ab}.
\ee
More surprisingly, all 16 central charges $Z_{ab}$ appear also as
zero
modes, namely as the 10 fluxes of the self-dual three-form
$db^+$ through the 3-cycles of
$T^5$, the 5 winding numbers of the scalar $Y$, and the 5-flux of its
dual $*dY$.  Thus these charges are in one-to-one correspondence with
the odd homology of $T^5$, which naturally forms a spinor
representation of $SO(5,5,\Z)$. To complete the space-time
supersymmetry
algebra, one also has to use the world-brane supercharges
$\cQ^{a\a}$,
which act on the zero-modes as
\ba
\{\cQ^\a_a,S^\b_b\} \is \e^{\a\b} Z_{ab}, \nonu
\{\cQ^\a_a,S^{\dot\b}_b\} \is \omega_{ab} \pslash^{\a\dot\b}.
\label{first}
\ea
{}From these equations it follows that $\cQ^\a_a$ contains a
zero-mode
contribution $Z^{ab}S_{b\a} + \pslash^{\a\dot\b} S_{a\dot\b}$
in addition to the non-zero-modes that we have been considering up to
now. Then, from the world-brane supersymmetry algebra (\ref{left})
one
deduces that $\cP^0 1^{ab}+\cP^m\LL\Gamma_m^{ab}$ also contains a
zero
mode contribution $\hf(p_i^2 1^{ab} +(Z^\dagger Z)^{ab})$.

We can now use this to re-analyse the BPS conditions on the five
brane. In $d=6$ we have the condition that on physical states the
total space-like momentum on the five-brane vanishes,
$\cP^m\LL=0$.\footnote{This can be seen as a generalisation of the
$L_0-\Lbar_0=0$ condition in light-cone string theory.} This implies
that the oscillator contribution $P^m\LL$ and the zero-mode
contribution $K^m\LL={1\over 8}\tr(\G^m Z^\dagger Z)$ cancel, so that
we
derive that $P\LL^m = - K\LL^m$. Similarly one finds that
$P\RR^m=-K\RR^m$. The BPS mass formula now follows by imposing the
five-brane mass-shell condition $\cP^0 =p_+p_-$. This gives
$m\bps^2={1\over 4}\tr(Z^\dagger Z)+2|K|$, where $|K|=|K\LL|=|K\RR|$.
Thus the $SO(5,5)$ vector $(P\LL,P\RR)$ satisfies $P\LL^2-P\RR^2=0$,
and hence is indeed a null-vector.

The degeneracy formula follows now immediately by specializing the
general formula (\ref{dimsum}) to the case $\hat{P}^2=0$. So the
number of BPS states is given by $D(N_K)$, where $N_K$ is the number
of
times that the primitive vector $\hat{K}$ fits in $K$, and $D(N)$ is
the
degeneracy at level $N$ of the standard superstring partition
function
\be
\label{TTT}
\sum_N D(N) q^N=(16)^2\prod_n{\left({1+q^n\over 1-q^n}\right)^8}.
\ee
This degeneracy formula is the unique U-duality invariant extension
of
the results obtained in \cite{sen,vafa6d} from the counting of string
and D-brane BPS states. Note that pure string or D-brane
configurations carry at most 8 charges that transform as an $SO(4,4)$
vector $(q\,{}\LL,q\, {}\RR)$, and their degeneracy is $D(\hf(q\,
{}\LL^2\! -q\, {}\RR^2))$. The above result shows that for more
general bound states between string and D-branes, the degeneracy is
obtained by generalizing the T-duality invariant $\hf(q\, {}\LL^2\!
-q\, {}\RR^2)$ to the greatest common divisor of the ten components
of
the $SO(5,5)$ vector $(K\LL,K\RR)$ defined in (\ref{relations}).

\newsubsection{BPS Spectrum and Black Hole Entropy in $D=5$.}

It is straightforward to extend the six-dimensional results to $d=5$
and the BPS states that respect only $1/8$ of the space-time
supersymmetries.

The 5 anti-symmetric tensor fields in 6 dimensions can be decomposed
into 5 fields $B\LL^m$ with self dual field strength and 5 fields
$B\RR^m$ with anti-self dual field strength. Together
$(B\LL^m,B\RR^m)$ form a vector of $SO(5,5,\Z)$.  After further
compactification to $d=5$ these anti-symmetric tensor fields produce
$10$ additional gauge fields, with 10 corresponding charges
$(W^m\LL,W^m\RR)$. Together with the Kaluza-Klein momentum $p$ and
the
16 charges contained in $Z^{ab}$ this gives a total of $27$
charges. These charges combine into one irreducible representation of
the $E_{6(6)}(\Z)$ U-duality group: an $8\times 8$
pseudo-real, anti-symmetric and traceless matrix ${\cal Z}$
\cite{cremmer}.  This matrix can be expressed in terms of the
$SO(5,5)$
vector $(W\LL,W\RR)$, scalar $p$ and spinor $Z$ as
\be
\cZ = \twomatrixd {p +W\LL\! \cdot\G}{Z}{Z^\dagger}{-p+W\RR\!
\cdot\G}.
\ee

By dimensional arguments, the extremal black holes that carry these
charges
are expected to have a non-zero entropy that is the square root of a
cubic expression in ${\cal Z}$.
There is one unique cubic $E_{6(6)}$
invariant, namely $\tr\cZ^3$. In the above normalisation, this
leads to the following prediction for the U-duality invariant entropy
formula
\ba
S(\cZ) \is 2\pi \Bigl[\hf p (W\LL^2 -W\RR^2)+ {\textstyle {1\over 8}}
W\LL\! \cdot\tr(\G Z^\dagger Z)- {\textstyle{1\over 8}}W\RR\!\cdot
\tr(\G
Z Z^\dagger)\Bigr]^{1/2} \nonumber\\[2.5mm]
\is 2\pi
\Bigl[{1\over 2p}(|p W\LL \!- K\LL|^2 -|pW\RR\! -K\RR|^2)\Bigr]^{1/2}
\label{E6}
\ea
with $K$ as defined in (\ref{relations}).

How can such a result be derived from the five-brane?  Because we are
considering 1/8 BPS states in space-time, in this case we will only
have to impose the left-moving BPS conditions. So in terms of the
self-dual string one expects to get multiple string states on the
five-brane with right-moving oscillators. Note however, that this is
only possible when the momentum $(P^m\LL,P^m\RR)$ is no longer a null
vector, but satisfies $P^2>0$. Also, we need to represent the
additional charges $(W\LL^m,W\RR^m)$ and the Kaluza-Klein momentum
$p$
on the Hilbert states of the five-brane. The interpretation of the
vector charges $(W\LL^m,W\RR^m)$ is roughly that they indicate the
winding number of the five-brane around the extra compactification
circle. In this way the Kaluza-Klein momentum $p$ is fed into the
five-brane and this modifies the level matching relations on the
five-brane to $\cP\LL^m= pW\LL^m$ and similarly for the
right-movers.\footnote{The analogous statement in string theory is
that
$L_0-\Lbar_0$ no longer vanishes when the space-like direction of the
light-cone $(x^+,x^-)$ plane is taken to be compact. Instead it
equals
$p w$ with $p$ and $w$ the momentum and winding number in that
direction.} Now following the same steps as before one finds that the
total left-moving momentum of the multiple string states is
\be
\label{pisw}
P\LL^m=p W\LL^m - K\LL^m
\ee
and similarly for $P\RR^m$.

Even though we are able to represent all
27 charges, only the $SO(5,5,\Z)$ subgroup of the U-duality group
$E_{6(6)}(\Z)$ is a manifest symmetry of the five-brane. The full
U-duality is hidden in the light-cone construction, which has extra
subtleties because we work on a ``light-cone cylinder.'' Because of
this, our construction is in fact most straightforward for $p=1$.
In this case the counting of the states as explained above
equation (\ref{asymp}) goes through without change, and
leads for the statistical entropy $S = \log(\dim \cH_P)$ to
the following result
\be
S = 2\pi \sqrt{\half (W-K)^2}.
\ee
Comparing to the expected $E_{6(6)}$-invariant result (\ref{E6}) we
see that for $p\neq 1$ an extra factor $1/p$ enters. In the
light-cone
gauge constructions that we have been using, this factor naturally
follows from the relative normalisation of the time-coordinates on
the world-sheet of the self-dual string and space-time. In this way
we
find that the five-brane description of the BPS states exhibits a
maximal symmetry under the five-dimensional U-duality group.  The
fact that we furthermore reproduce the expected black hole entropy is
a
strong indication that the five-brane also gives an exhaustive
description of the BPS spectrum in $d=5$.  It would be interesting to
find a general covariant derivation of this result, starting from
a formulation as in \cite{bergshoefetal}, since this could
make the full $E_{6(6)}$-invariance manifest from the start.

\newsubsection{Concluding remarks}

We have shown that at least down to $5$ dimensions the string
formulation of the five-brane theory gives a unified description of
all BPS states that is invariant under the U-duality group.
The general quantisation of five-dimensional extended objects is
still an unsolved problem, but we see convincing evidence that the
``BPS
quantisation'' of the five-brane reduces it effectively to a type II
string, since the momenta $k^m$ on the world-brane are aligned with
the central charge $K^m$. U-duality permutes these various strings,
similar to the action of $SL(2,\Z)$ on the $(p,q)$ strings in type
IIB string theory in ten dimensions. The correspondence with the type
II
string also shows immediately that our description for the spectrum
of 1/4 BPS states is fully Lorentz invariant. This is however
somewhat
less obvious for the 1/8 BPS states.

We can also consider compactifications of M-theory on other manifolds
than tori, such as orbifolds \cite{orbifolds}.  Particular cases that
we have analyzed (see \cite{inprep}) are $K3\times S^1$ and $T^4
\times (S^1/\Z_2)$.  In both of these compactifications the
five-brane
can be shown to behave as a heterotic string. In the latter case, the
geometric explanation is that the worldbrane geometry of the
five-brane
must also be of the form of $K3\times S^1$. This result can be used
to
give a new explanation from M-theory of the various heterotic
string/string dualities \cite{duffetal}.  Furthermore, the counting
of
BPS states can be done in similar manner as in this letter
\cite{inprep}, and is in agreement with the black hole entropy
formula
derived in \cite{strovaf}.

Finally, let us comment on the generalization to four dimensions.
Here one gets one extra electric charge $p^\prime$ from the
Kaluza-Klein momentum, which gives a total of $28$ charges.  But a  
general state in $d=4$ can carry $28$ magnetic and $28$ electric  
charges that combine into the $56$ of the $E_{7(7)}(\Z)$ U-duality
group.  In this case the entropy is expected to scale with the square
root of a quartic expression in the central charge, which is now a
general complex anti-symmetric $8 \times 8$ matrix.  The unique
quartic invariant is the so-called diamond function $\diamondsuit$.
In
\cite{kallosh} this was used to conjecture that the macroscopic
entropy in four dimensions equals $S=2\pi \sqrt \diamondsuit$.  For
the states obtained from the five-brane, this entropy formula can be  
expressed in terms of $SO(5,5,\Z)$ representations and takes the form
\be
S = 2\pi  \sqrt{{p'\over 2p}(pW-K)^2}.
\ee
We believe that via a straightforward generalization of the above
procedure one should be able to reproduce this result. This would
give a complementary derivation of the recently obtained results
\cite{4dblackholes}. It seems, however, that to obtain a
fully $E_{7(7)}$-invariant description from the five-brane, new
ingredients may be needed. Possibilities are the inclusion of D-brane
configurations on the world-volume \cite{douglas}, or of bound states
of several five-branes \cite{mem,intmem}, which would lead to
non-abelian extensions of the self-dual string theory considered
here.

\bigskip

{\noindent \sc Acknowledgements}

We would like to thank S. Ferrara, C. Kounnas, W. Lerche, R.
Minasian, and C. Vafa for interesting discussions and helpful
comments.  This
research is partly supported by a Pionier Fellowship of NWO, a
Fellowship of the Royal Dutch Academy of Sciences (K.N.A.W.), the
Packard Foundation and the A.P. Sloan Foundation.

\renewcommand{\Large}{\large}

\end{document}